\documentclass[journal,twoside]{IEEEtran}
\pdfoutput=1
\usepackage{amsmath,amsfonts}
\usepackage{array}
\usepackage[caption=false,font=normalsize,labelfont=sf,textfont=sf]{subfig}
\usepackage{textcomp}
\usepackage{stfloats}
\usepackage{url}
\usepackage{verbatim}
\usepackage{graphicx}
\usepackage{siunitx}

\usepackage{cite}

\begin{document}
%
\title{Mitigating Motion Sickness with Optimization-based Motion Planning}
%
%
%

\author{Yanggu~Zheng$^{1}$,
        Barys~Shyrokau$^{1}$,
        and~Tamas~Keviczky$^{2}$,
\thanks{The research leading to these results has received funding from the European Union Horizon 2020 Framework Program, Marie Sklodowska-Curie actions, under grant agreement no. 872907.}
\thanks{$^{1}$Yanggu Zheng and Barys Shyrokau are with the Department of Cognitive Robotics, Faculty of Mechanical, Maritime and Materials Engineering,
        Delft University of Technology, 2628CD, The Netherlands
        {\tt\small y.zheng-2@tudelft.nl, b.shyrokau@tudelft.nl}).}
\thanks{$^{2}$Tamas Keviczky is with Delft Center for Systems and Control, Faculty of Mechanical, Maritime and Materials Engineering,
        Delft University of Technology, 2628CD, The Netherlands
        {\tt\small t.keviczky@tudelft.nl}}}

%
%

\markboth{arXiv preprint, January 2023}
{Zheng \MakeLowercase{\textit{et al.}}: Mitigating Motion Sickness with Optimization-based Motion Planning}
%



\maketitle

\begin{abstract}
The acceptance of automated driving is under the potential threat of motion sickness. It hinders the passengers' willingness to perform secondary activities. In order to mitigate motion sickness in automated vehicles, we propose an optimization-based motion planning algorithm that minimizes the distribution of acceleration energy within the frequency range that is found to be the most nauseogenic. The algorithm is formulated into integral and receding-horizon variants and compared with a commonly used alternative approach aiming to minimize accelerations in general. The proposed approach can reduce frequency-weighted acceleration by up to 11.3\% compared with not considering the frequency sensitivity for the price of reduced overall acceleration comfort. Our simulation studies also reveal a loss of performance by the receding-horizon approach over the integral approach when varying the preview time and nominal sampling time. The computation time of the receding-horizon planner is around or below the real-time threshold when using a longer sampling time but without causing significant performance loss. We also present the results of experiments conducted to measure the performance of human drivers on a public road section that the simulated scenario is actually based on. The proposed method can achieve a 19\% improvement in general acceleration comfort or a 32\% reduction in squared motion sickness dose value over the best-performing participant. The results demonstrate considerable potential for improving motion comfort and mitigating motion sickness using our approach in automated vehicles.
\end{abstract}


%
\IEEEpeerreviewmaketitle

\section{Introduction}
\IEEEPARstart{A}{utomated} driving is facing a paradoxical situation created by motion sickness. Drivers will be freed from the driving task and allowed to engage in secondary activities such as reading and writing. This could potentially boost the productivity of society as a whole. However, when drivers become passengers, they lose the ability to anticipate vehicle motion and consequently become more susceptible to motion sickness. The symptoms, ranging from slight dizziness and fatigue to nausea and vomiting \cite{bos2005motion}, could hinder them in performing secondary activities. Therefore, motion sickness is considered as a threat that could potentially undermine a large portion of the societal benefits promised by automated driving. \par
Significant research effort has been directed towards understanding, modeling, and ultimately overcoming motion sickness. Motion sickness is estimated to influence almost half of the adult passengers in road transport \cite{golding2006predicting}. It has been suspected to be a result of sensory conflict \cite{reason1978motion,oman1990motion,bos1998modelling}, although some consider it to be related to the loss of postural stability \cite{riccio1991ecological}. Early studies focused primarily on the vertical movements that are most significant on surface vessels. The concept of motion sickness dose value (MSDV) was proposed for vertical oscillations, indicating that the incidence of vomiting is strongly correlated to a value equivalent to the root-mean-square acceleration times the square-root of the duration \cite{lawther1987prediction}. \par 
In road transport, however, passengers are more often subjected to lateral and longitudinal disturbances. A series of studies were conducted focusing on the occurrence of motion sickness in public road transport \cite{turner1999motionA,turner1999motionB,turner1999motionC}, where a strong correlation between motion sickness and low-frequency lateral accelerations was observed. The horizontal translational oscillations at around 0.2 Hz are found to be the most nauseogenic according to experiments on human subjects \cite{golding2001motion}. Further research suggests that between 0.0315 and 0.2 Hz, the lateral oscillation is comparably provoking \cite{donohew2004motion}. The relative importance between longitudinal and lateral motion in provoking sickness is not validated but it was recommended that the same weighting could be used \cite{donohew2004motion}. The comfort threshold for fore-aft, lateral and vertical vibrations was studied but for a higher frequency range than what is interesting for motion sickness research \cite{morioka2006magnitude}. Meanwhile, no significant variance in the subjects' frequency sensitivity from 0.15 to 0.5 Hz has been observed in terms of group-average motion sickness index but the individual differences were strong\cite{irmak2021individual}. Nevertheless, the findings in these studies agree that low-frequency acceleration is the major contributing factor to motion sickness. Therefore, mitigation of motion sickness relies on the reduction of low-frequency accelerations.\par
The motion planner on an automated vehicle plays a critical role in the vehicle's comfort level. Given the perception inputs on available driving space and other road users, a motion planer essentially generates the desired path and velocity profile that low-level tracking controllers should then follow. Early studies aim at improving general motion comfort in a qualitative manner by ensuring the continuity of the motion. There are works on constructing the motion with curve components that feature continuous curvature. This guarantees a bounded jerk if the vehicle travels at a constant speed and is often achieved with variations of clothoids \cite{silva2018clothoid,zhang2019hybrid}. Some studies exploit the smoothness of parametric curves, e.g., splines and Bezier curves, to plan a path or velocity profile \cite{qian2016motion,artunedo2019real,lattarulo2020real,cao2022implementing}. These approaches ensure that the higher-order time derivatives of the motion profile are bounded. The parametric curves can also be incorporated into an optimization-based framework where the placement of control points is optimized. In addition, state lattices and motion primitives have been used by some authors to improve smoothness under a sampling- or optimization-based planning framework \cite{bottasso2008path,mcnaughton2011motion,mischinger2018towards}. Most of the studies mentioned above do not explicitly consider motion sickness or accelerations in general and therefore have limited applicability in the dynamic driving scenarios that automated vehicles encounter. Although there has been an attempt to analyze the frequency components in the resulting acceleration patterns arising from the use of different transition curves \cite{siddiqi2021ergonomic}. \par
Some studies exploit optimization techniques to reduce acceleration. A comfort-oriented planner was proposed for car-like mobile robots, where the motion is planned by minimizing a cost function concerning travel time, accelerations, and clearance from an obstacle \cite{shin2018kinodynamic}. This study also mentioned the trade-off between time and comfort in motion planning. The trade-off problem in real traffic is investigated for a roundabout scenario where naturalistic driving data is analyzed \cite{zheng2021comfort}. Furthermore, an optimal planning framework has been proposed, which considers the possibility of reducing lateral accelerations with active roll motion \cite{zheng20223dop}. Some recent works paid more attention to mitigating motion sickness by considering the frequency sensitivity discussed above. Explicit minimization of MSDV in addition to travel time has been explored in \cite{htike2021fundamentals}, where the vehicle motion is formulated in curvilinear coordinates and details were provided on the distribution of acceleration energy in the frequency domain and on the trade-off between comfort and travel time \cite{htike2021fundamentals}. An alternative approach to the problem was proposed using a time-domain planning algorithm by including a second-order high-pass filter in the objective function \cite{li2021mitigating}. The weighting is done in the frequency domain after performing the Fourier transform of the acceleration signals.\par
We believe automated driving holds more potential in mitigating motion sickness than human drivers in a similar way it does in other aspects. Automated vehicles are believed to have superior sensing capability while humans are better at reasoning and inferring \cite{schoettle2017sensor} when trying to understand the surrounding environment. Regarding safety, a lower crash rate has been observed from automated vehicles that have a safe driver onboard \cite{teoh2017rage}. It has been argued that equal safety performance from automated vehicles would not be sufficient to greatly promote its acceptance \cite{liu2020self} and it could be generalized to other aspects, too. In terms of planning and control, we believe automated vehicles are able to consistently perform motion patterns that help mitigate motion sickness, while human drivers may find it demanding or even challenging to do so. This opinion currently lacks solid support from actual research data on human drivers' capability of planning vehicle motion. A quantitative understanding of human drivers' performance in planning vehicle motion can be beneficial for developing motion planning algorithms and promoting the acceptance of automated vehicles.\par

The major contributions of this paper include:
\begin{itemize}
    \item Design of an optimization-based motion planning algorithm that can be used to mitigate motion sickness or improve general acceleration comfort.
    \item Detailed explanations of calculating frequency-weighted accelerations using a band-pass filter. This is different from the high-pass filter used in \cite{li2021mitigating} where the high-frequency fluctuations are expected to be filtered out by the limited actuation bandwidth. The effectiveness of frequency weighting is further demonstrated by comparing it to using an alternative objective, where all accelerations are penalized equally.
    \item A receding-horizon formulation of the optimization-based motion planning problem that is closer to real-world deployment. The real-time capability of the receding-horizon planner has been examined for different combinations of parameters.
    \item An initial experimental effort to measure human driving data for the purpose of evaluating motion planning algorithms. The best-performing participant was selected as an example for demonstrating the potential of AVs in improving motion comfort.
\end{itemize}

The rest of the paper is organized as follows. Section II introduces the proposed motion planning scheme in detail, including the formulation of the optimal planning problem and the implementation of frequency weighting. Section III describes the scenario chosen for testing the algorithm and Section IV explains the experimental setup for measuring human drivers on real roads. Extensive results and discussions are presented in Section V. Finally, Section VI concludes the paper by summarizing the findings and limitations of this study and suggesting potential directions for future research.
\section{Optimization-based Motion Planning Algorithms}

\noindent This section introduces the underlying components that construct an optimization problem reflecting the goal of mitigating motion sickness in automated vehicles. The vehicle motion is defined within the lane space using the lateral position relative to the lane center, in addition to vehicle velocity. The relative lateral position and velocity are the decision variables that enable the evaluation of the objective function that should be minimized. The optimization problem is formulated for an integral approach and a receding-horizon approach. The former represents the ideal performance while the latter reflects a level that is achievable in practice.\par

\subsection{Motion Definition}
\noindent The proposed motion planner assumes that the vehicle drives on well-paved roads with clear lane marks so that the information about the available driving space along the road is clearly defined by the perception systems. The vehicle trajectory is constructed through a series of waypoints. Along the center of the lane where the vehicle is driving, a string of stations is distributed with a nominal interval $d_\text{nom}$. To locate a waypoint with regards to its corresponding station, we first define a local lateral axis at the station perpendicular to the normal driving direction. Then the location of the waypoint $P_k$ is determined by its lateral position $y_k$. The station is placed at $y_k=0$, i.e., at the lane center. With a positive $y_k$, the vehicle deviates to the left-hand side of the lane. Connecting the waypoints results in a spatial trajectory while assigning a velocity to each waypoint further defines the trajectory in time. This allows for calculating the accelerations of the vehicle when the steps below are followed. First, the Euclidean distance between waypoint $k$ and $k+1$ is given by:
\begin{equation}
    {d_k} = \sqrt {{{\left( {{X_{k + 1}} - {X_k}} \right)}^2} + {{\left( {{Y_{k + 1}} - {Y_k}} \right)}^2}}
\end{equation}
\noindent Assuming that the vehicle has constant longitudinal acceleration when traveling this distance, we could determine the longitudinal acceleration given the velocity at waypoint $k$ and $k+1$ with:
\begin{equation}
    {a_{x,k}} = \frac{{v_{k + 1}^2 - v_k^2}}{{2{d_k}}}
\end{equation}
\noindent Further assuming that the vehicle heading at waypoint $k$, denoted by ${\vec h}_k$, points from waypoint $k$ to waypoint $k+1$:
\begin{equation}
    {{\vec h}_k} = \left( {\begin{array}{*{20}{c}}
    {{X_{k + 1}}}\\
    {{Y_{k + 1}}}
    \end{array}} \right) - \left( {\begin{array}{*{20}{c}}
    {{X_k}}\\
    {{Y_k}}
    \end{array}} \right)
\end{equation}
\noindent The change of heading can be calculated as the angle between ${{\vec h}_k}$ and ${{\vec h}_{k+1}}$:
\begin{equation}
    {\psi _k} = \frac{{{{\vec h}_k} \cdot {{\vec h}_{k + 1}}}}{{\left| {{{\vec h}_k}} \right|\left| {{{\vec h}_{k + 1}}} \right|}}
\end{equation}
\noindent The curvature of the vehicle path in this section is given by:
\begin{equation}
    \kappa_k = \frac{{\Delta {\psi _k}}}{{{d_k}}}
\end{equation}
\noindent The lateral acceleration is then approximated using the average speed: 
\begin{equation}
    {a_{y,k}} = {\left( {\frac{1}{2}{v_{k + 1}} + \frac{1}{2}{v_k}} \right)^2}{\kappa_k}
\end{equation}
\noindent Here, the accelerations are calculated purely from the waypoint locations and velocities without considering vehicle dynamics. In this formulation, we only constrain the lateral position of the waypoints according to the lane width and the velocities according to speed limits. Because the main goal of this study is to explore how to effectively incorporate the frequency sensitivity in motion sickness into an optimization-based motion planning scheme, and to demonstrate how time efficiency and motion comfort can be balanced. Therefore, we try to design here a general motion plan in the form of a spatiotemporal trajectory that receives minimal influence from the choice of prediction model and constraints. The accurate tracking of the desired trajectory is expected to be handled by the low-level tracking controllers that incorporate individual differences in vehicle dynamics. In addition, including a vehicle dynamics model in the optimization problem increases the computational complexity. If the computation time from the current formulation is sufficiently short and the resulting motion plan would show signs of infeasibility, it is possible to add a low-level loop consisting of a vehicle dynamics model and a tracking controller, resembling the approach proposed in \cite{kuwata2009real}.\par
\subsection{Objectives}
\noindent The primary objective of the proposed method is to minimize the incidence of motion sickness indicated by frequency-weighted accelerations. Nevertheless, the motion planner should also take into account the factor of time efficiency in the optimization. This prevents the planner from driving unnecessarily slow and hence ensures a driving behavior that is acceptable to the passengers and other road users. We define the objective of the optimization as a weighted sum of motion sickness and time efficiency:
\begin{equation}\label{eqn:obj_ms}
    J_\text{MS} = D_\text{MS} + WT
\end{equation}
\noindent Where $D_{MS}$ is a measure of motion sickness, $T$ is a measure of time efficiency, and $W$ is the weighting factor on time efficiency. A larger $W$ means a shorter travel time is preferred over a less sickening ride. A viable range of $W$ should be determined beforehand through simulation while the user could be given the freedom to adjust it during the trip according to personal preferences. The time efficiency is described by the total duration of the planned motion. Under the assumption of constant longitudinal acceleration between two adjacent waypoints, travel time at a given step can be found as:
\begin{equation}\label{eqn:dt}
    \Delta t_k = \frac{2d_{k}}{{v_{k + 1}} + {v_k}}
\end{equation}
\noindent The total duration is simply the sum of the travel time of all the steps:
\begin{equation}\label{eqn:T}
    T = \sum\limits_k {\Delta {t_k}}
\end{equation}
\noindent Measuring motion sickness, on the other hand, is less straightforward. We adopt the following form as is used in \cite{shin2018kinodynamic, zheng2021comfort}:
\begin{equation}\label{eqn:d_ms1}
    {D_\text{MS, Motion}} = \sum\limits_{k = 1}^N {\left( {a_{xf,k}^2 + a_{yf,k}^2} \right)\Delta {t_k}}
\end{equation}
\noindent Where, $a_{xw}$ and $a_{yw}$ are the frequency-weighted longitudinal and lateral accelerations, respectively. The total energy of the frequency-weighted accelerations is used to quantify the comfort level related to motion sickness. This is effectively the square of MSDV for longitudinal and lateral motion proposed in \cite{htike2020minimisation}:

\begin{equation}\label{eqn:msdv}
    \text{MSDV} = {\left( {\int_0^T {\left( {a_{xw}^2 + a_{yw}^2} \right)} dt} \right)^{1/2}}
\end{equation}

\noindent The benefit of using squared MSDV instead is that the penalization on discomfort is amplified by itself. This choice may not directly impact the resulting trade-off given that $W$ is varied through a wide range. Rather, it influences the way that $W$ regulates the balance between time and comfort. With the squared MSDV, $W$ has to be increased more sharply to further promote aggressive driving behavior. This could be more friendly to the users when they are given the option of $W$ especially when they are inexperienced. \par
In order to apply frequency-weighting on accelerations, we implement two separate band-pass filters, incorporating the different frequency sensitivities in the longitudinal and lateral directions. Each band-pass filter can be expressed as a transfer function constructed from a low-pass filter and a high-pass filter:
\begin{equation}\label{eqn:bp_tf}
    H\left( s \right) = \frac{{{a_{\text{fil}}}\left( s \right)}}{{{a_{\text{act}}}\left( s \right)}} = \frac{1}{{{\tau _1}s + 1}}\frac{s}{{{\tau _2}s + 1}}
\end{equation}
\noindent Where $\tau_1$ and $\tau_2$ are the time constants corresponding to the desired cut-off frequencies of the band-pass filter. The transfer function has an equivalent continuous-time state-space formulation of:
\begin{equation}\label{eqn:bp_ss_cont}
\begin{array}{l}
\dot x = Ax + B{a_{\text{act}}}\\
{a_\text{fil}} = Cx
\end{array}
\end{equation}
\noindent The state-space matrices are calculated as:
\begin{equation}
\begin{array}{l}
A = \left( {\begin{array}{*{20}{c}}
{ - {\tau _1}^{ - 1} - {\tau _1}^{ - 1}}&1\\
{ - {\tau _1}^{ - 1}{\tau _2}^{ - 1}}&0
\end{array}} \right)\\
B = \left( {\begin{array}{*{20}{c}}
{{\tau _1}^{ - 1}{\tau _2}^{ - 1}}\\
0
\end{array}} \right)\\
C = \left( {\begin{array}{*{20}{c}}
1&0
\end{array}} \right)
\end{array}
\end{equation}
\noindent Given a time step of $\Delta t$ and assuming zero-order hold for the input, the state-space model can be discretized as:
\begin{equation}
\begin{array}{l}
{x_{k + 1}} = {A_d}{x_k} + {B_d}{a_{\text{act},k}}\\
{a_{\text{fil},k}} = {C_d}{x_k}
\end{array}
\end{equation}
\noindent Where,
\begin{equation}
\begin{array}{l}
{A_d} = {e^{A\Delta t}}\\
{B_d} = {A^{ - 1}}\left( {{A_d} - I} \right)B\\
{C_d} = C
\end{array}
\end{equation}
\noindent Given the current filter states and assuming the actual acceleration to be constant through a time period $\Delta t$, it is possible to calculate the filtered acceleration at the end of the period using the method described above. This is useful to find the frequency-weighted acceleration when the vehicle travels between two consecutive waypoints. Because the step travel time, $\Delta t_k$ varies per waypoint, the matrix exponential $e^{A\Delta t_k}$ is different for every time step. Diagonalization is a commonly used technique in calculating matrix exponentials. Instead of diagonalizing $A\Delta t_k$ for every given $\Delta t_k$, we first diagonalize matrix $A$:
\begin{equation}
    \Omega = {P^{ - 1}}AP
\end{equation}
\noindent The matrix exponential is then equivalent to:
\begin{equation}
    {e^{A\Delta t}} = P{e^{\Omega \Delta t}}{P^{ - 1}}
\end{equation}
\noindent The matrix exponential of diagonal matrix $\Omega \Delta t$ is simply:
\begin{equation}
    {e^{\Omega \Delta t}} = \left( {\begin{array}{*{20}{c}}
    {{e^{{\omega_{11}}\Delta t}}}&0\\
    0&{{e^{\omega_{22}\Delta t}}}
    \end{array}} \right)
\end{equation}
\noindent Because matrix $P$ is invariant of $\Delta t$, it can be pre-computed without the need to repeat the computation in every evaluation step. Using these steps, we reduce the complexity of computation from a matrix exponential to the exponential of two real numbers on top of basic matrix multiplications with a size of 2-by-2. This allows us to efficiently determine the value of $D_\text{MS,Motion}$ given a time-stamped acceleration sequence. However, the output of a band-pass filter accumulated through the travel time is not equal to the squared MSDV because the filter has its temporal dynamics. Both the internal states of the filter and the output require extra time to dissipate and converge to zero after the input is removed. This process needs to be taken into account as the output in this period contains a part of the information describing the nauseogenicity of the acceleration input. Neglecting it would allow the optimal solution to exploit the filter dynamics by commanding unreasonably high acceleration towards the end of the planning horizon. Hence, we calculate the output of the filter given zero input for a period of 30 seconds at a sampling time of 0.2s and include this amount in the evaluation of the planned motion:
\begin{equation}\label{eqn:d_ms2}
    {D_\text{MS, Tail}} = \sum\limits_{k = N+1}^{N+N_\text{Tail}} {\left( {a_{xf,k}^2 + a_{yf,k}^2} \right)\Delta {t_k}}
\end{equation}
The choice of 30 seconds is based on the choice of $\tau_1$ for lateral accelerations. The total cost term reflecting motion sickness is:
\begin{equation}\label{eqn:d_ms_total}
    {D_\text{MS}} ={D_\text{MS, Motion}} + {D_\text{MS, Tail}}
\end{equation}
\noindent 
To demonstrate the effectiveness of the proposed formulation in targeting a specific frequency range in accelerations, we add an alternative objective where the frequency dependency is neglected. This variant is further referred to as minimal-acceleration planning or MA in short. It minimizes a cost function of similar form to \eqref{eqn:obj_ms} although the term $D_\text{MS}$ is replaced with $D_\text{MA}$, a measure of general acceleration comfort:
\begin{equation}\label{eqn:d_ac}
    {D_\text{MA, Motion}} = \sum\limits_{k = 1}^N {\left( {a_{x,k}^2 + a_{y,k}^2} \right)\Delta {t_k}}
\end{equation}
The rest of its implementation is identical to the proposed sickness-mitigating planning method (further referred to as MS).
\subsection{Integral Approach}
\noindent The integral approach here refers to planning for the entire driving scenario by solving a single large-scale optimization problem in the following form:
\begin{equation}\label{eqn:ocp}
    \begin{array}{rc}
        \text{min:} & J(\textbf{X}) \\
        \text{where:} & \textbf{X} = \left[ {{y_1} \ldots {y_N},{v_1} \ldots {v_N}} \right] \\
        \text{s.t.:} & {y_{\min }} \le {{y_1} \ldots {y_N}} \le {y_{\max }}\\
        & {v_{\min }} \le {{v_1} \ldots {v_N}} \le {v_{\max }}\\
    \end{array}
\end{equation}

\noindent where $N$ is the total number of stations for the scenario. The choice of $J$ depends on the purpose of the planner. It contains either $D_\text{MS}$ for mitigating motion sickness or $D_\text{MA}$ for minimizing overall acceleration. The process of evaluating $J$ using the decision variables $\textbf{X}$ given a road profile has been shown above. In the case of the integral approach, the motion planner is given all the information about the maneuvering scenario and produces an optimal motion plan in one go. This approach requires detailed and highly accurate information and could be difficult to realize in practice. In case of unpredictable events from other road users, the motion plan may have to be abandoned in order to respond. It nevertheless provides a benchmark that represents the ideal performance that a motion planner could achieve. It can be used to evaluate other planning methods that aim to improve computational efficiency.\par

\subsection{Receding-horizon Approach}

\noindent In the receding-horizon approach, the planner solves an optimization problem repeatedly over a receding horizon, concerning motion predictions for only a limited distance ahead:
\begin{equation}\label{eqn:ocp_rh}
    \begin{array}{rc}
        \text{min:} & J(\textbf{X}_\text{RH}) \\
        \text{where:} & \textbf{X}_\text{RH} = \left[ {{y_1} \ldots {y_{Np}},{v_1} \ldots {v_{Np}}} \right] \\
        \text{s.t.:} & {y_{\min }} \le {{y_1} \ldots {y_{Np}}} \le {y_{\max }}\\
        & {v_{\min }} \le {{v_1} \ldots {v_{Np}}} \le {v_{\max }}\\
    \end{array}
\end{equation}
\noindent where $N_p$ represents the number of stations distributed within the preview distance $D_p$. It concerns a shorter distance ahead that is visible to the vehicle instead of being given the full knowledge of the driving scenario. The rest of the formulation is identical to the integral approach. This approach is more applicable in the real world as it receives up-to-date information about a limited distance ahead. After reaching the first waypoint planned previously, new information is processed to update the motion plan. In this way, the vehicle could also respond to unforeseeable changes in the traffic situation. The optimization problem to be solved at each step has significantly lower complexity and is easier to solve on automotive-level hardware, potentially enabling real-time operation. Although the limited preview range may impact the optimality of the planned motion when compared to the integral approach that gives the theoretically optimal solution.\par 
The preview distance is expected to be the most influential factor and needs to be chosen with care. However, a fixed preview distance cannot achieve a good balance between having good resolution and keeping a low complexity when the vehicle speed is allowed to vary in a wide range. A long preview distance is needed to ensure safety when driving on motorways while a shorter preview distance is enough for driving in built-up areas. In order to account for varying velocities, we propose a speed-dependent preview distance for the receding-horizon planner. Instead of a fixed value, we instead calculate the preview distance using the vehicle's velocity multiplied by a chosen preview time $T_p$, i.e., $D_{p, k}=v_k T_p$. The distance is divided equally into $N_p$ intervals to find reference stations. The $N_p$ defined in \eqref{eqn:ocp_rh} is further referred to as the planning horizon. Consequently, the nominal sampling time $T_s$ is $T_p/N_p$. Because in reality, the travel time between two consecutive waypoints depends on the velocity in that segment. We investigate the performance corresponding to different combinations of $T_p$ and $N_p$ as listed in Table \ref{tab:rh_params}. We choose $T_p=$ \SI{3}{\second} as the minimum for safety concerns. It is realistic to consider a case where the vehicle has to reduce its speed from \SI{100}{\kilo\meter\per\hour} to a full stop. With a preview time of \SI{3}{\second}, the vehicle has approximately \SI{83}{\meter} to complete the deceleration process, leading to an average longitudinal deceleration of approximately \SI{4.7}{\meter\per\square\second}. Sustained deceleration with such a level of aggressiveness will not be considered as comfortable and a preview time shorter than this may even be a threat to safety. On the other hand, we chose the longest preview time of \SI{5}{\second}. This is based on the observed computation time when combining $T_p=$\SI{5}{\second} and $T_s=$\SI{0.1}{\second}. The simulation then becomes extremely time-consuming.\par

\begin{table}
    \centering
    \begin{tabular}{ccc}
        $T_p$ [s] & $N_p$ [-] & $T_s$ [s]\\
        \hline
        3 & 30 & 0.1\\
        3 & 15 & 0.2\\
        3 & 6  & 0.5\\
        \hline
        4 & 40 & 0.1\\
        4 & 20 & 0.2\\
        4 & 8  & 0.5\\
        \hline
        5 & 50 & 0.1\\
        5 & 25 & 0.2\\
        5 & 10 & 0.5
    \end{tabular}
    \caption{Preview Parameters of the receding-horizon planner}
    \label{tab:rh_params}
\end{table}

\section{Evaluation Method}
\subsection{Simulation Scenario}
\begin{figure}
    \centering
    \includegraphics[width=\linewidth]{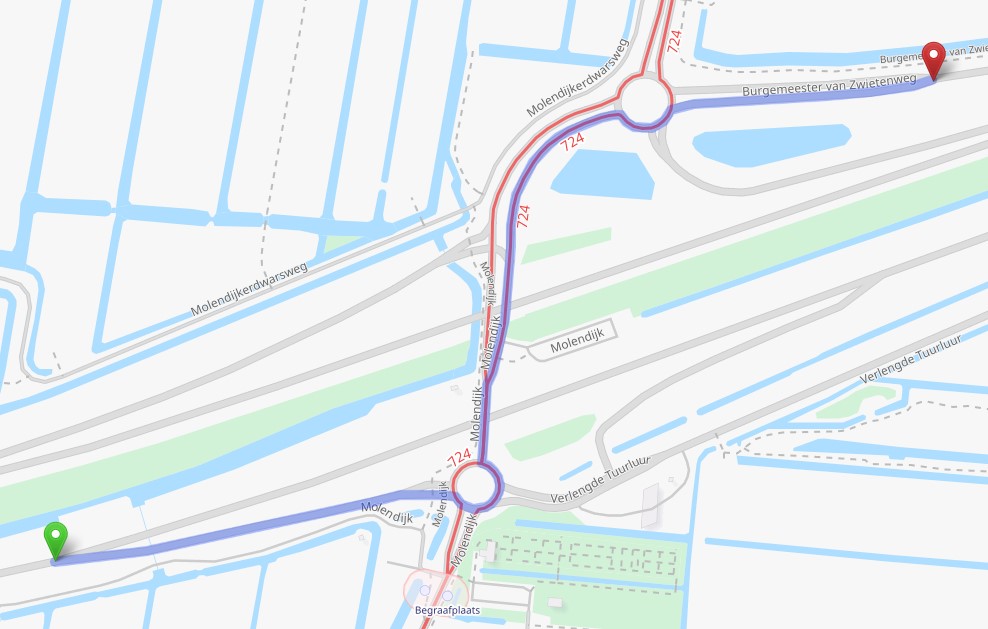}
    \caption{The test route chosen for simulation evaluation of the motion planners and the human driving experiment (Image source: Open Street Map).}
    \label{fig:route}
\end{figure}

\noindent A real-world road section (see Fig.\ref{fig:route}) in the Netherlands has been chosen for the purpose of testing the proposed motion planner. The section starts from the exit ramp of motorway A12 (52°03'52.0"N 4°49'04.0"E) and ends on a distribution road (52°04'03.8"N 4°49'39.1"E). The total driving distance is approximately \SI{920}{\meter}. The vehicle needs to navigate through 2 roundabouts and make 11 turns (4 left-handed, 7 right-handed). Furthermore, the vehicle starts at \SI{100}{\kilo\meter\per\hour} before decelerating and entering the first turn and needs to accelerate back to \SI{80}{\kilo\meter\per\hour} after the last turn. Each straight sector for the significant speed changes has a length of approximately \SI{200}{\meter}. Excluding these two parts, the vehicle needs to negotiate a turn roughly every \SI{50}{\meter} or \SI{4}{\second}, making the scenario challenging for human drivers and motion planners alike. Using a public road section with such high maneuvering intensity for the simulation purpose is therefore considered meaningful and revealing. The road section is modeled from satellite images as a sequence of line segments and circular arcs. \par

\subsection{Human Driving Data}
\noindent We believe that comparing motion planning algorithms and human drivers is highly valuable for the development and promotion of automated driving. However, to the best of our knowledge, a human driver baseline is not available in the literature for our purpose of investigation. Therefore, we conducted a real-world experiment to collect the data for this purpose. The Human Research Ethics Committee (HREC) of TU Delft has reviewed and approved the experiment procedures under application 2405. In the current pilot phase of the experiment, a total of 6 participants have volunteered to drive the test vehicle through the route described above. All of them are male, in the age range of 25-30, and have held valid driving licenses for a minimum of 3 years. The participants were informed of the purpose of the experiment beforehand so that they could possibly adjust their driving style. The number of participants is relatively small due to resource constraints. Hence we select the participant with the best performance for comparison purposes. This would at least be representative of an above-average human driver. To ensure minimal influence from other road users, all test drives were performed between 7 and 9 pm when the traffic is minimal. The experiment was conducted in July so sufficient ambient lighting was available when the tests took place. All participants were given enough time and driving distance to familiarize themselves with the test vehicle. This ensures that they performed test runs with confidence in their ability to control the vehicle. \par
A Hyundai Tucson hybrid was used for the experiment. The automatic transmission simplified the learning process for the participants. The test vehicle was instrumented with a VBOX RLVBSS100 for GPS-based position and velocity measurement, and an XSENS MT-28A for orientations and linear accelerations. The GPS sensor is programmed to automatically start and stop measuring using a coordinate-based trigger. The IMU provides no option for an external trigger and has to be operated manually when the vehicle passes certain 'landmark' objects. As part of the route crosses underneath the motorway, the connection to satellites is lost for a period of \SI{15}{\second} on average. In post-processing, we first naively infer this part of vehicle trajectory by linearly interpolating between the last correct reading of coordinates before the loss of connection and the first correct reading after re-establishing the connection. To combine data from both sources, we utilize the linear Kalman filter scheme. Two separate filters are used here, one for the $X$ direction and the other for $Y$. Essentially, the Kalman filter predicts a 1-D kinematic process where the position is a result of double-integrating acceleration. The accelerations measured by the IMU are originally in the vehicle's local frame and must be transformed into the global frame using the yaw angle. In each filter, the GPS position is used as the measurement while the projected acceleration is used as the input. Because the GPS velocity is computed by differentiating the position, this piece of information is not used in the filter apart from when initializing the internal states. The covariance matrices are initialized in favor of the IMU readings. It should be emphasized that the reconstructed motion using this method underestimates the actual accelerations because the intermediate section consists of an S-turn but is interpolated linearly due to a lack of information. We attempted to re-initialize the covariance matrices at the beginning of the interpolated section to further reduce the filter's reliance on positional data. Unfortunately, this caused a discontinuity in the estimated trajectory. Therefore, we accept the fact that the overall acceleration is underestimated because at least we do not claim an exaggerated advantage of our method over human drivers. \par

\section{Results and Discussions}

\subsection{Integral planning}

\begin{figure}
    \centering
    \includegraphics[trim={20pt 25pt 40pt 25pt}, clip, width=\linewidth]{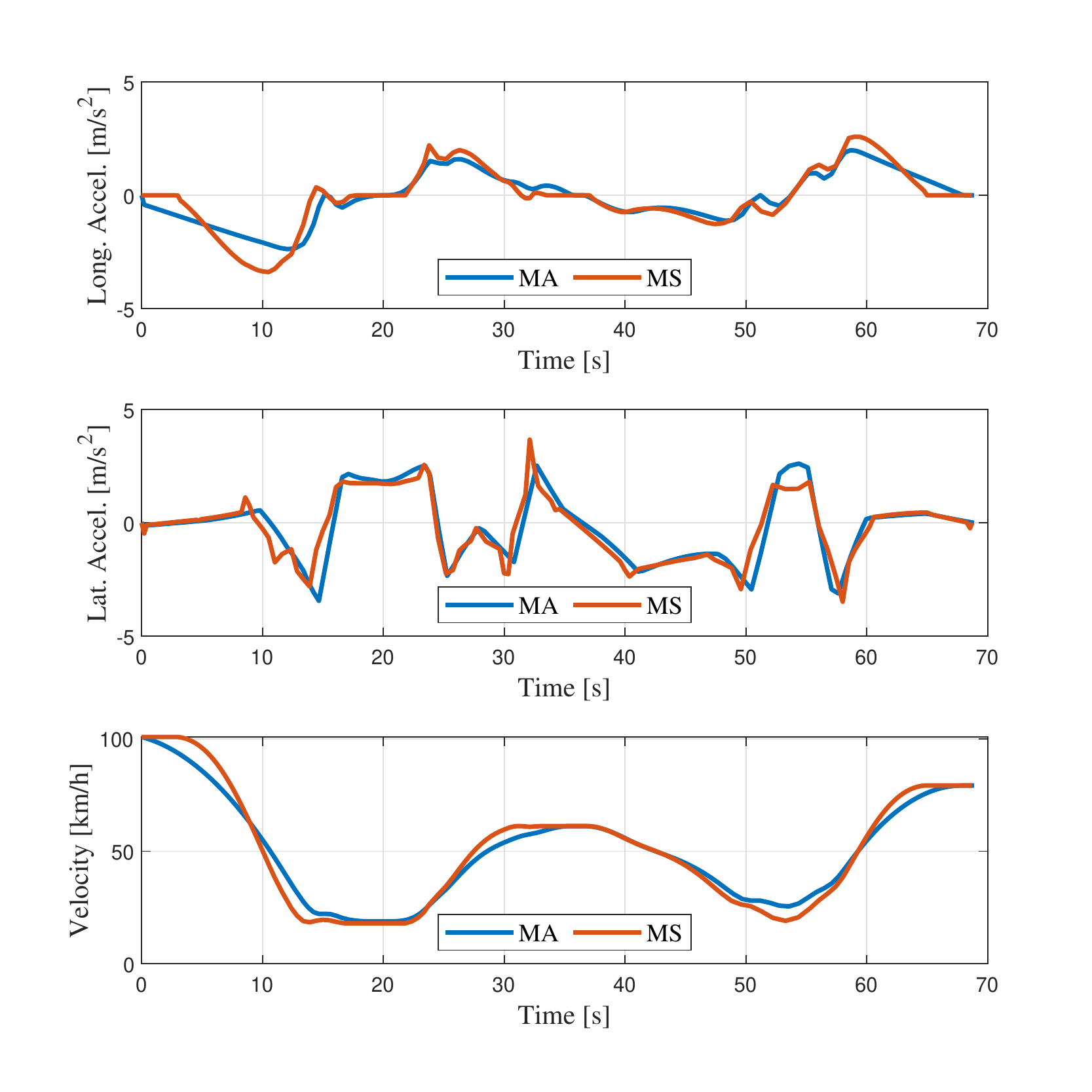}
    \caption{Comparison of motion profiles between planner variations aimed at motion sickness mitigation and acceleration minimization.}
    \label{fig:ac_ms}
\end{figure}

\noindent The comparison between MS and MA planners using the integral approach can be found in Fig. \ref{fig:ac_ms}. Both motion profiles have a travel time of approximately \SI{69}{\second}. This value is chosen according to the fastest human driver run that we have recorded. The MS planner initiates the first deceleration later and with a higher peak magnitude than the MA planner. The entry velocity to the first right-hand turn is significantly lower (at around \SI{12}{\second}), leading to a lower lateral acceleration through the first roundabout. At the intermediate part of the road, the MS planner commands a swifter acceleration to the speed limit, exhibiting higher peak lateral acceleration through the turns. Then upon entering the second roundabout, the MS planner again adopts a lower speed for the turning part and then accelerates more aggressively to reach the speed limit of \SI{80}{\kilo\meter\per\hour}. In general, the MS planner is more aggressive in changing vehicle speed while taking sharper corners more gently. This is partly an effect of applying separate filters to the longitudinal and lateral accelerations.\par

\subsection{Varying parameters in receding-horizon planning}
\begin{figure} 
    \centering
    \includegraphics[trim={20pt 0pt 40pt 10pt}, clip, width=\linewidth]{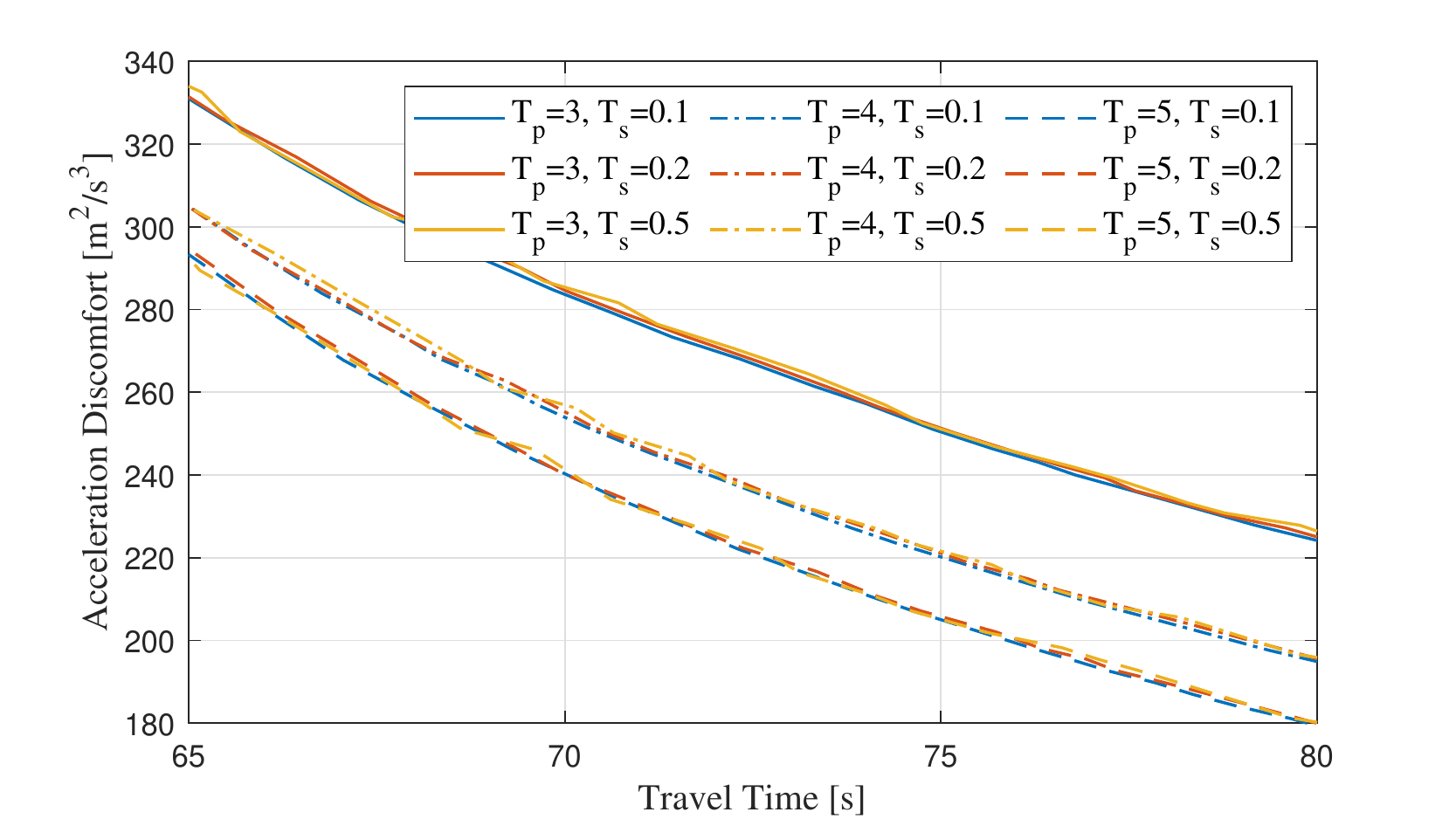}
    \caption{Comparison of acceleration comfort and time efficiency performance of receding-horizon planner variations using different preview times and sampling times.}
    \label{fig:dt_ac}
\end{figure}

\begin{figure}
    \centering
    \includegraphics[trim={20pt 20pt 40pt 20pt}, clip, width=\linewidth]{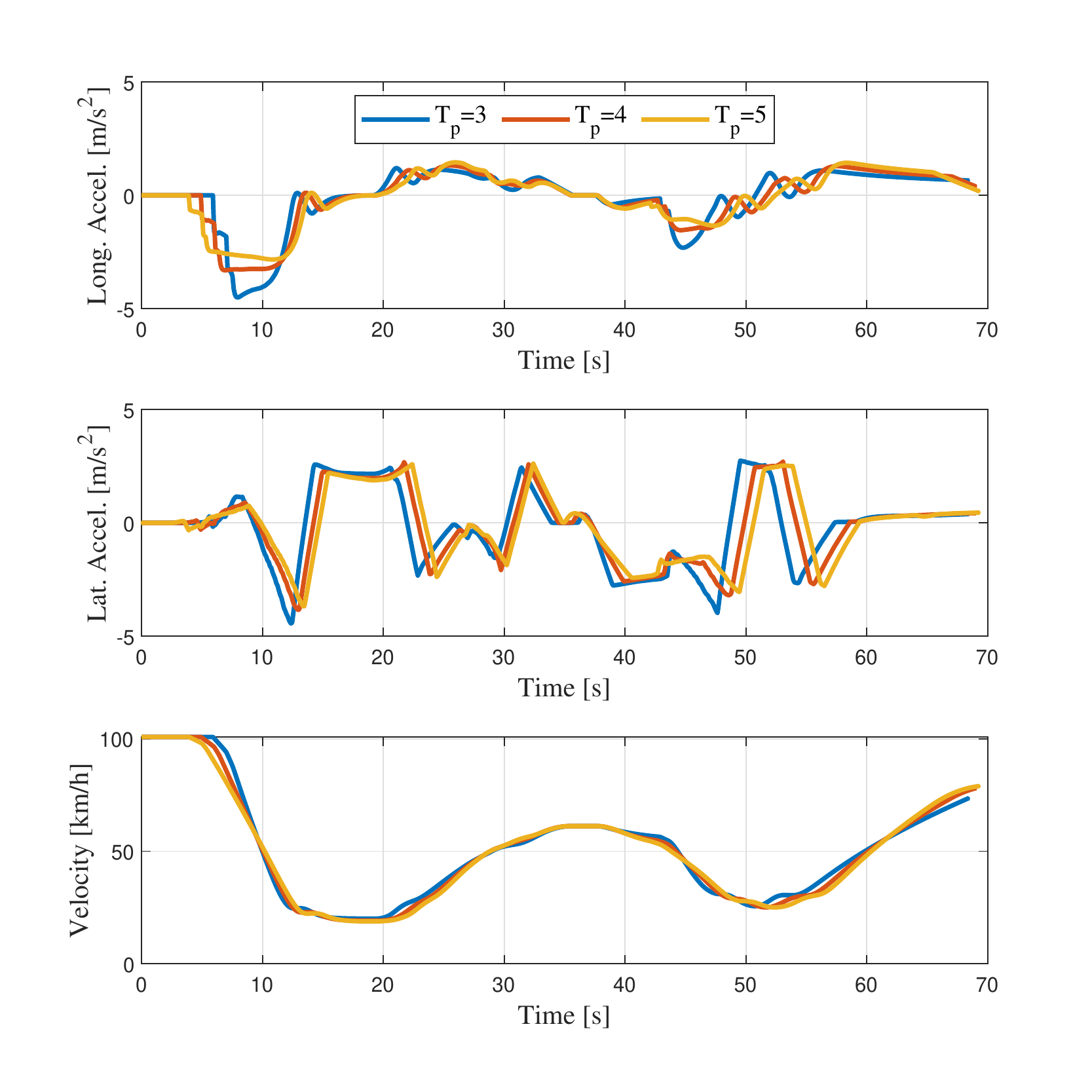}
    \caption{Comparison of motion profiles of the receding-horizon motion planner aimed at acceleration minimization using different preview times.}
    \label{fig:rh_ac}
\end{figure}

\begin{figure}
    \centering
    \includegraphics[trim={20pt 0pt 40pt 10pt}, clip, width=\linewidth]{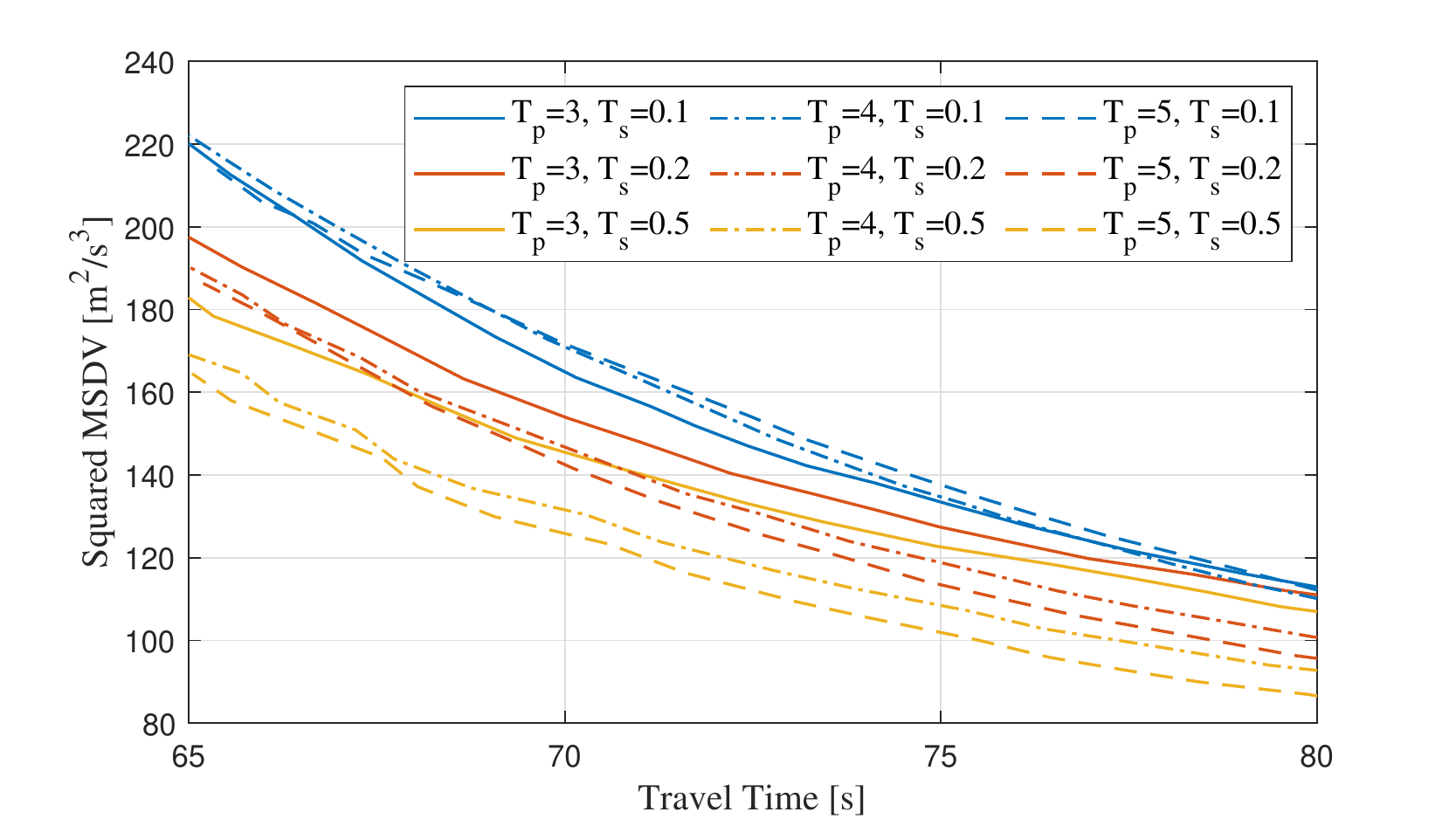}
    \caption{Comparison of nauseogenicity and time efficiency performance of receding-horizon planner variations using different preview times and sampling times.}
    \label{fig:dt_ms}
\end{figure}

\begin{figure}
    \centering
    \includegraphics[trim={20pt 20pt 40pt 20pt}, clip, width=\linewidth]{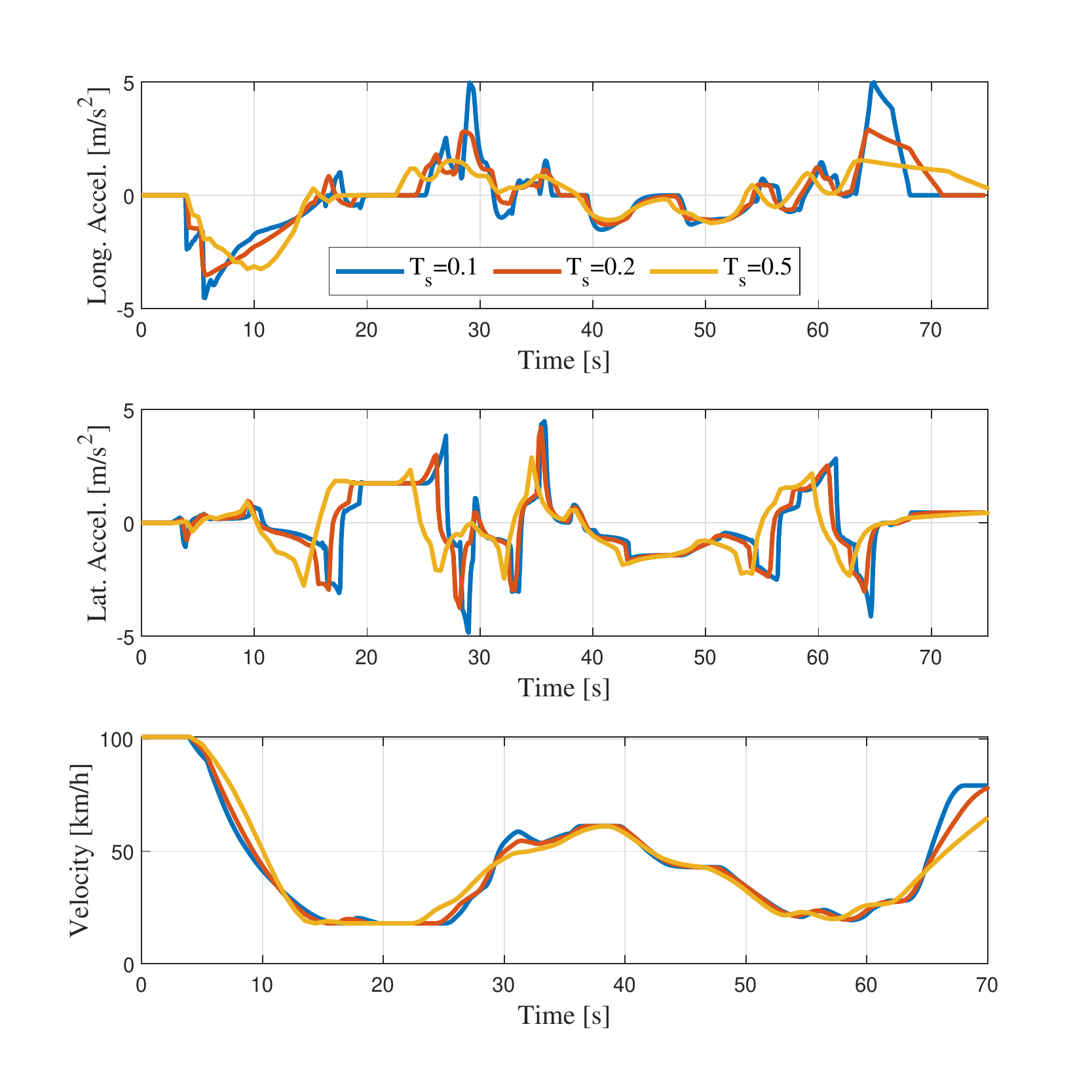}
    \caption{Comparison of motion profiles of the receding-horizon motion planner aimed at motion sickness mitigation using different sampling times.}
    \label{fig:rh_ms}
\end{figure}

\noindent When using the receding-horizon planner, the preview time $T_p$ and planning horizon $N_p$ (or the resulting nominal sampling time $T_s$) greatly influence the planning performance. One may naturally expect that a longer $T_p$ and larger $N_p$ would be beneficial. This is mostly true in the case of MA planning as Fig. \ref{fig:dt_ac} suggests. The loss of acceleration comfort is limited when using a smaller $N_p$ or longer $T_s$ up to \SI{0.5}{\second}. The major difference is caused by preview time as highlighted in Fig. \ref{fig:rh_ac}. All three motion profiles are with $T_s=0.2s$ and have a duration of \SI{69}{\second}. It is obvious from the first braking maneuver that a longer preview time leads to more gentle deceleration. At the first right-hand turn, a shorter preview time leads to more aggressive turning due to the less desirable positioning of the vehicle (i.e., not as far to the left). At around \SI{30}{\second}, the planner with a shorter preview time is more reluctant in speeding up between consecutive turns because the benefit of taking the turns slowly outweighs the time-saving effect when they are calculated for a shorter distance ahead. Similar observations can be made at the second roundabout. The MS planner behaves somewhat differently than expected when varying $T_p$ and $T_s$. The former has a similar effect on the planned motion as is observed with the MA planner, that a longer $T_p$ leads to better overall performance. The latter, however, causes a more significant difference that is acting in the opposite direction. Contrary to what is commonly expected, a longer $T_s$ has led to better overall performance when combined with a longer $T_p$ (Fig. \ref{fig:dt_ms}). This might be a consequence of using a discrete-time band-pass filter. When the step time between two waypoints $\Delta t_{k}$ is shorter, the planner is given the opportunity to command a very high initial acceleration without getting penalized due to the slow dynamics of the filter. Then because only the first step in this plan is actually applied and the rest is discarded, the resulting motion features more abrupt jumps in acceleration. This statement is supported by Fig. \ref{fig:rh_ms} where all three motion profiles are with $T_p=5s$ and have a duration of \SI{75}{\second}.\par

\subsection{Efficacy of frequency weighting}
\noindent It has been demonstrated by previous examples that incorporating frequency sensitivity in the optimization scheme leads to obvious changes in the shape of the accelerations. To quantify whether the features are reflected by quantitative measures, we collect and compare the value of $D_\text{MS}$. In the MS planners, the value is directly available by solving the optimization problem. In the MA planners, however, we need to obtain this by passing the time-stamped acceleration sequence through the band-pass filter. The long-tail effect is also taken into account here. Using the integral approach, the sickness-mitigating planner shows a squared MSDV 7.5-11.3\% lower than the minimal-acceleration planner across the range of reasonable travel time. At the same time, the MS planner causes an increase of approximately 8.3-11.7\% in acceleration discomfort over the MA planner. This observation suggests a potential conflict of interest between mitigating motion sickness and improving general motion comfort.\par

\begin{figure}
    \centering
    \includegraphics[trim={10pt 0pt 20pt 10pt}, clip, width=\linewidth]{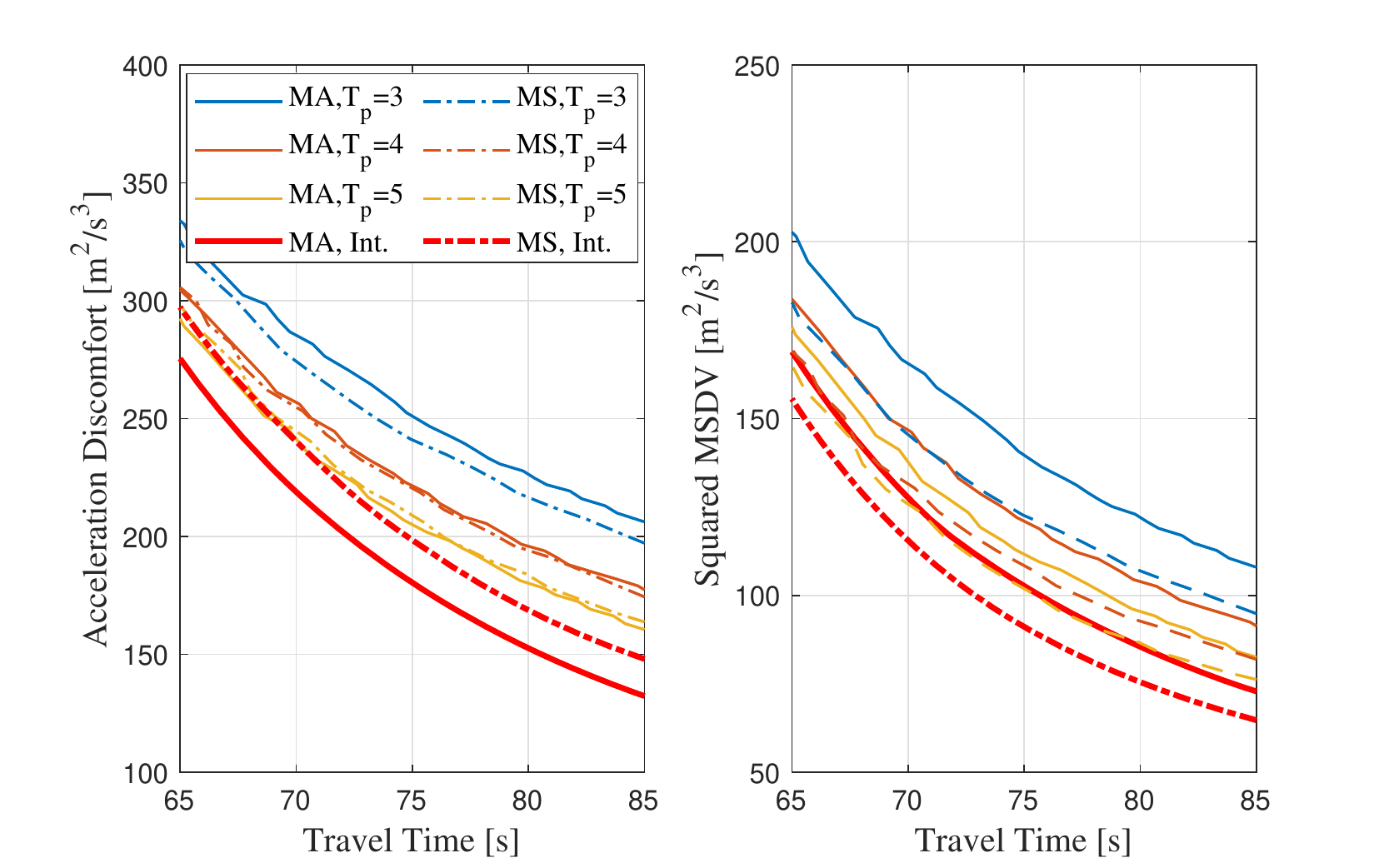}
    \caption{Demonstration of performance loss of the receding-horizon approach over the integral approach.}
    \label{fig:ms_ac_diff}
\end{figure}

\subsection{Acceleration magnitudes and feasibility}
\begin{figure}
    \centering
    \includegraphics[trim={10pt 0pt 20pt 10pt}, clip, width=\linewidth]{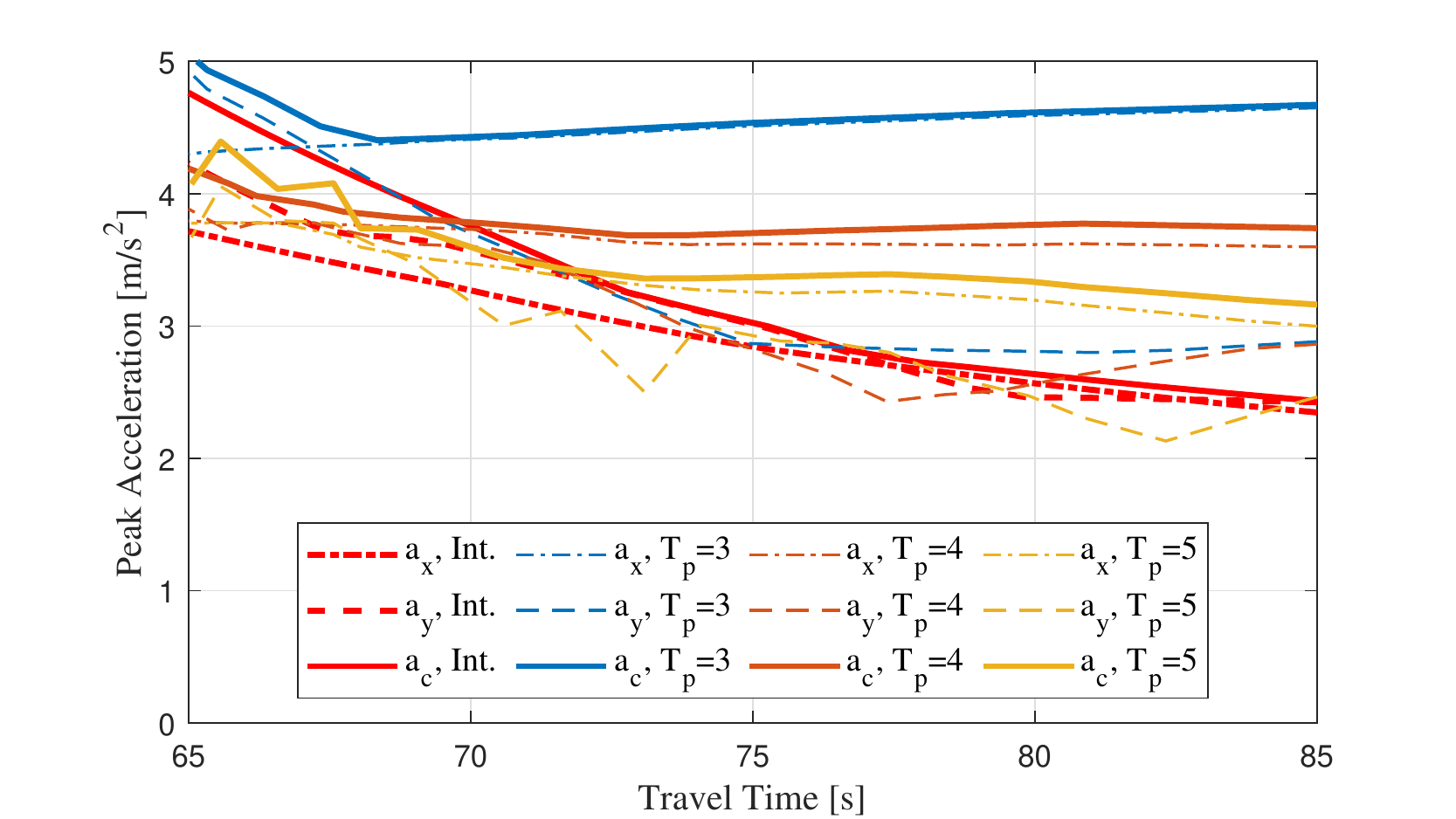}
    \caption{Maximum magnitude of longitudinal, lateral, and combined acceleration observed from the MS planner variants. The preview time is varied from 3s to 5s while the sampling time is fixed at 0.5s.}
    \label{fig:amax_ms}
\end{figure}

\begin{figure}
    \centering
    \includegraphics[trim={10pt 0pt 20pt 10pt}, clip, width=\linewidth]{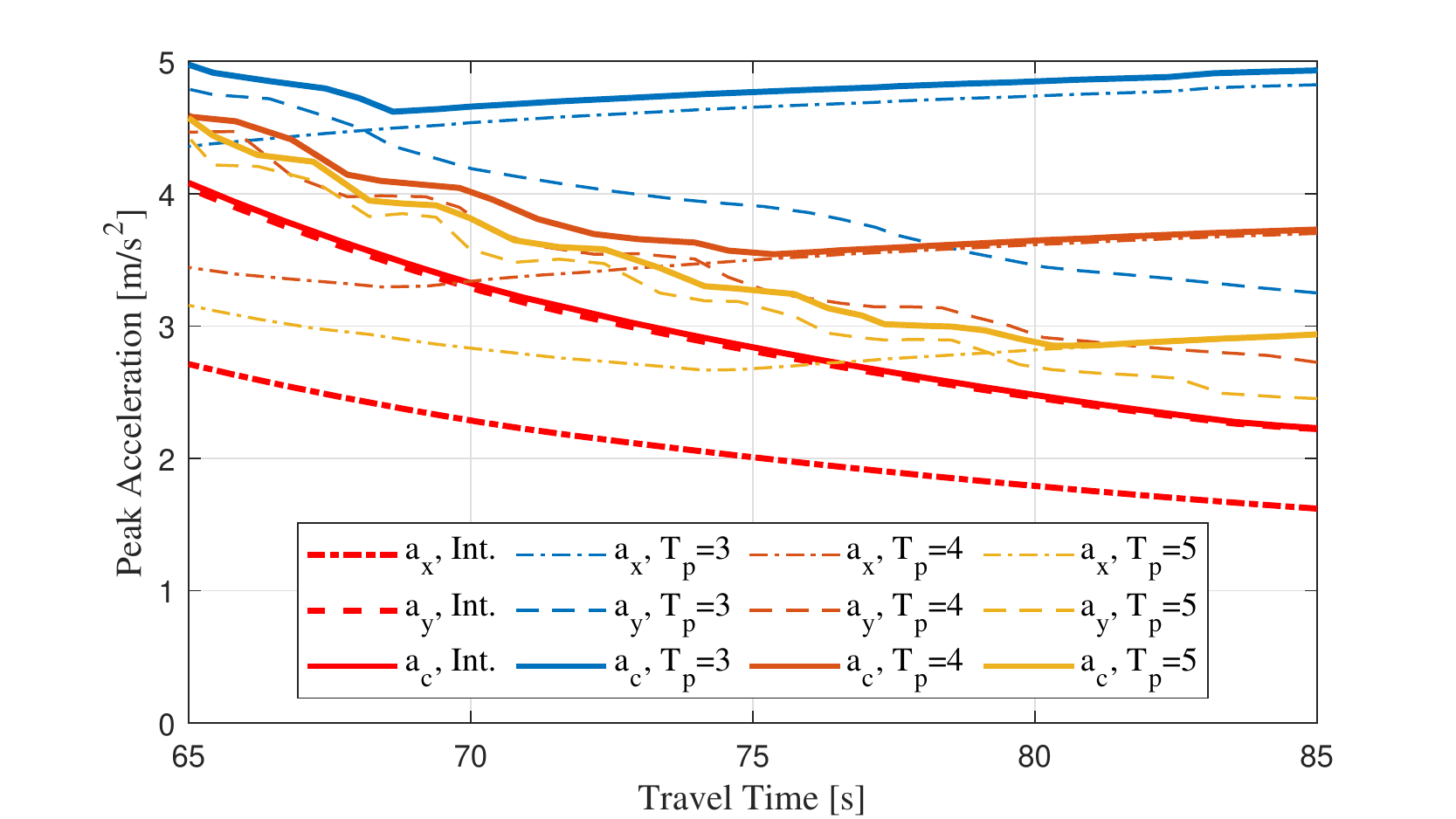}
    \caption{Maximum magnitude of longitudinal, lateral, and combined acceleration observed from the MA planner variants. The preview time is varied from 3s to 5s while the sampling time is fixed at 0.2s.}
    \label{fig:amax_ac}
\end{figure}

\noindent The optimization scheme does not impose any explicit constraint on the aggressiveness of the planned motion. The physical feasibility of the planned motion is verified by the peak acceleration magnitude. Fig. \ref{fig:amax_ms} shows the maximum absolute acceleration to be experienced by the vehicle with the MS planners,  while Fig. \ref{fig:amax_ac} shows that of the MA planners. The former exhibits slightly higher combined acceleration in general. The values observed here are well aligned with the findings in \cite{liu2017driving} where a peak acceleration of around \SI{6.0}{\meter\per\square\second} for urban driving is observed from naturalistic driving data. It can be seen from the receding-horizon approaches using either objective that the peak combined acceleration is closely following the peak longitudinal acceleration for more gentle driving, and switches to the peak lateral acceleration in between. When time efficiency receives a smaller weight, the vehicle reduces its speed so much as to achieve smaller lateral acceleration throughout the turns. A larger weight, on the other hand, causes the planner to prefer cornering faster to save time, consequently leading to a higher peak value in lateral acceleration.\par


\subsection{Computation Time}
\begin{figure}
    \centering
    \includegraphics[trim={10pt 0pt 20pt 10pt}, clip, width=\linewidth]{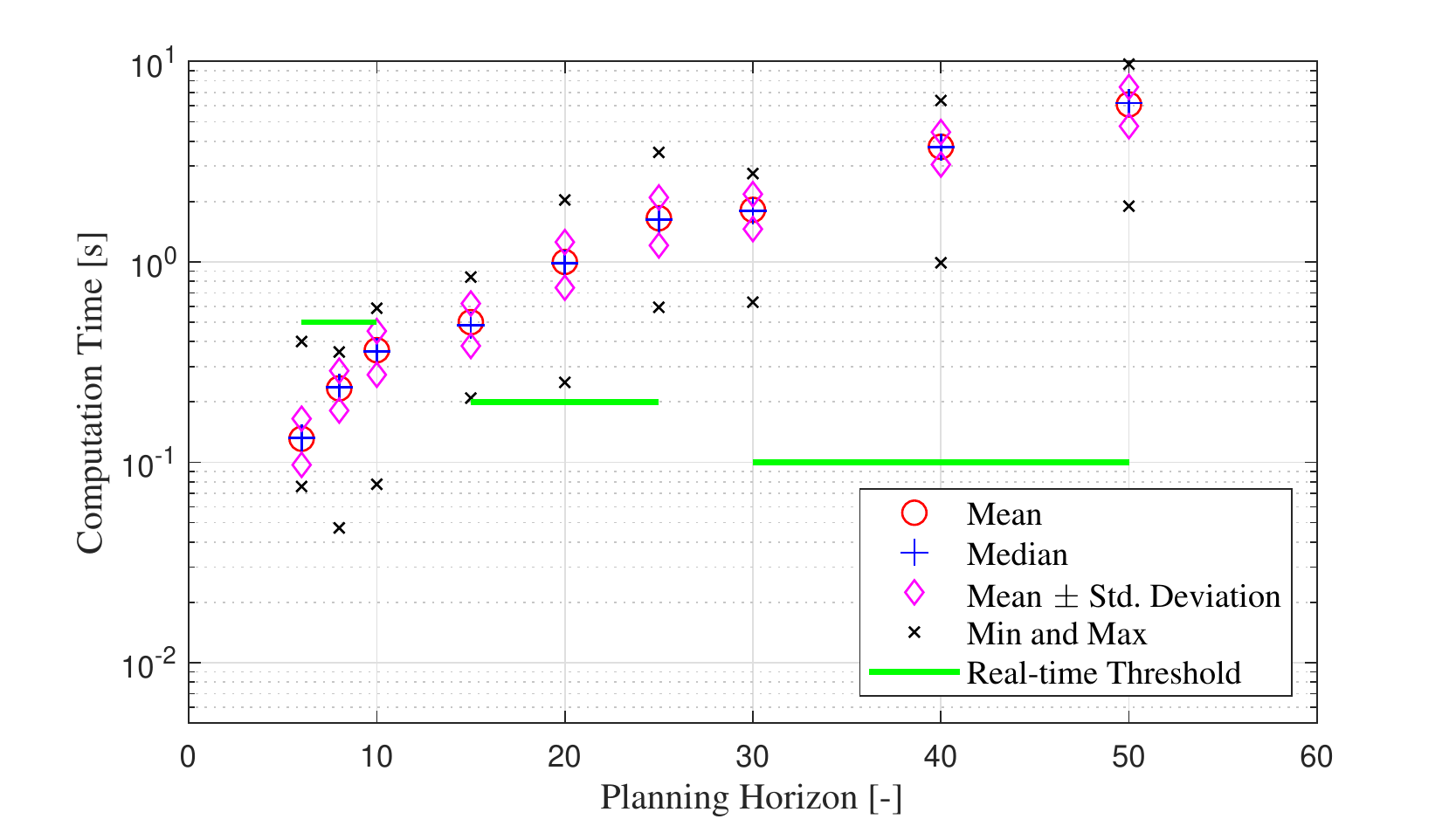}
    \caption{Computation time of receding-horizon MS planners using different preview times and sampling times.}
    \label{fig:tc_rh_ms}
\end{figure}

\begin{figure}
    \centering
    \includegraphics[trim={10pt 0pt 20pt 10pt}, clip, width=\linewidth]{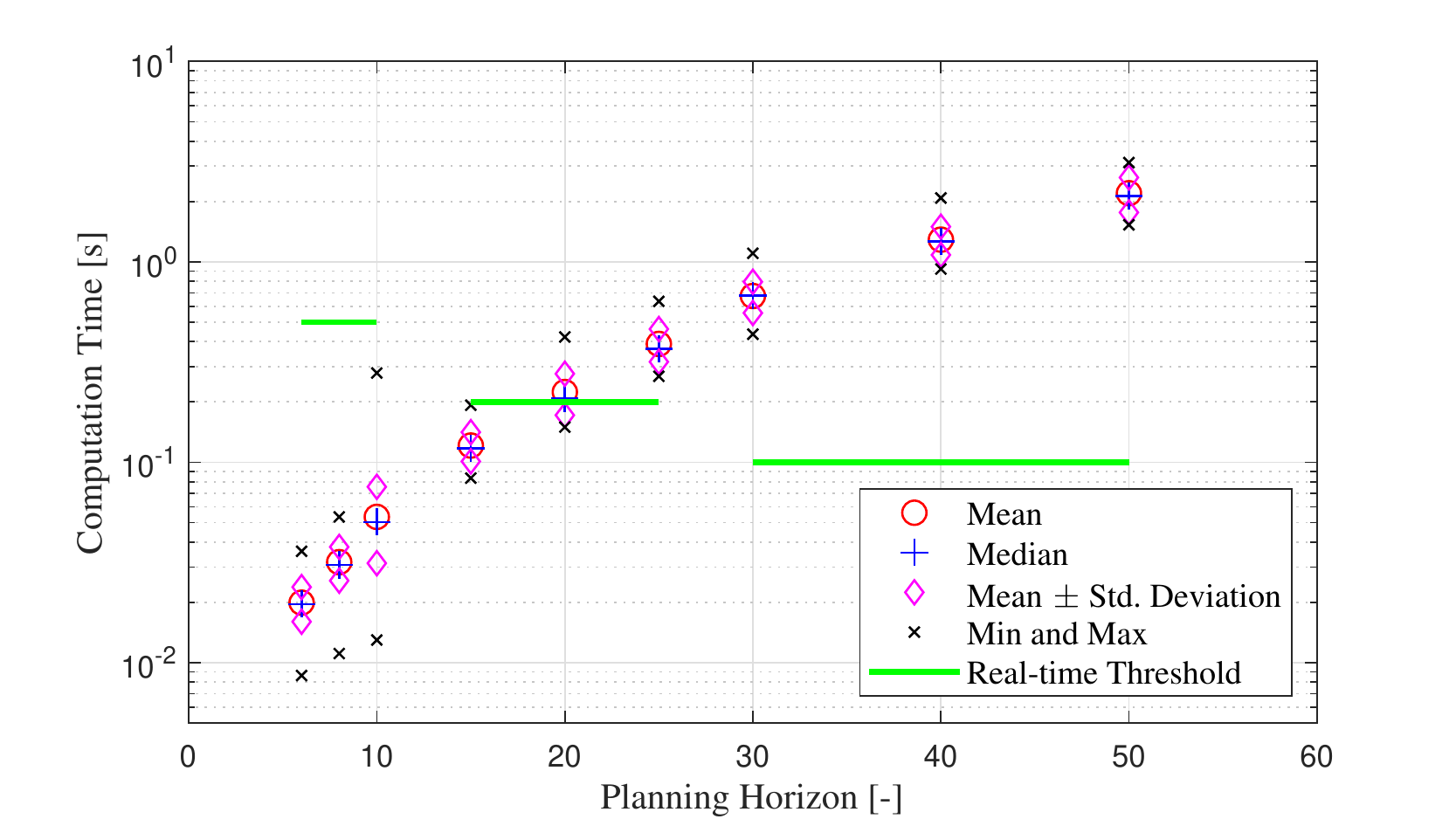}
    \caption{Computation time of receding-horizon MA planner using different preview times and sampling times.}
    \label{fig:tc_rh_ac}
\end{figure}

\noindent The receding-horizon approach is expected to have a lower computation complexity and to be capable of real-time implementation. We collected the computational time when the algorithm runs as MATLAB scripts on a desktop PC. The computation is limited to a single CPU core with a clock frequency of \SI{4.1}{\giga\hertz}. Fig. \ref{fig:tc_rh_ms} and \ref{fig:tc_rh_ac} show the variation in computation time with different $N_p$. With the same $T_p$, increasing $T_s$ from \SI{0.1}{\second} to \SI{0.2}{\second} reduces computation time by a factor of 5-6. An extra 5-fold reduction can be observed when $T_s$ increases to \SI{0.5}{\second}. With $T_s=$ \SI{0.5}{\second}, the MS planner is marginally capable of operating at $T_p=$\SI{5}{\second}. We expect that performing only a limited number of iterations and easing the termination criteria could guarantee that the peak computation time is below the real-time threshold. The algorithm may also be accelerated when running as compiled code instead of MATLAB script. The computation time of the MA planner is approximately 25\% to 30\% of the MS planner. This allows for a wider margin from the real-time threshold. Alternatively, it could be feasible to use the setting of $T_p=$ \SI{5}{\second} and $T_s=$ \SI{0.2}{\second} if the algorithm can indeed run faster after the modifications proposed above. The comparison between Fig. \ref{fig:tc_rh_ms} and \ref{fig:tc_rh_ac} highlights the complexity of incorporating the frequency sensitivity in motion sickness. It remains an open question whether the MS planner's effectiveness in mitigating motion sickness is worth the extra computation. This needs to be investigated further with experimental studies on human subjects.\par


\subsection{Comparison with Human Drivers}
\begin{figure}
    \centering
    \includegraphics[trim={20pt 25pt 40pt 20pt}, clip, width=\linewidth]{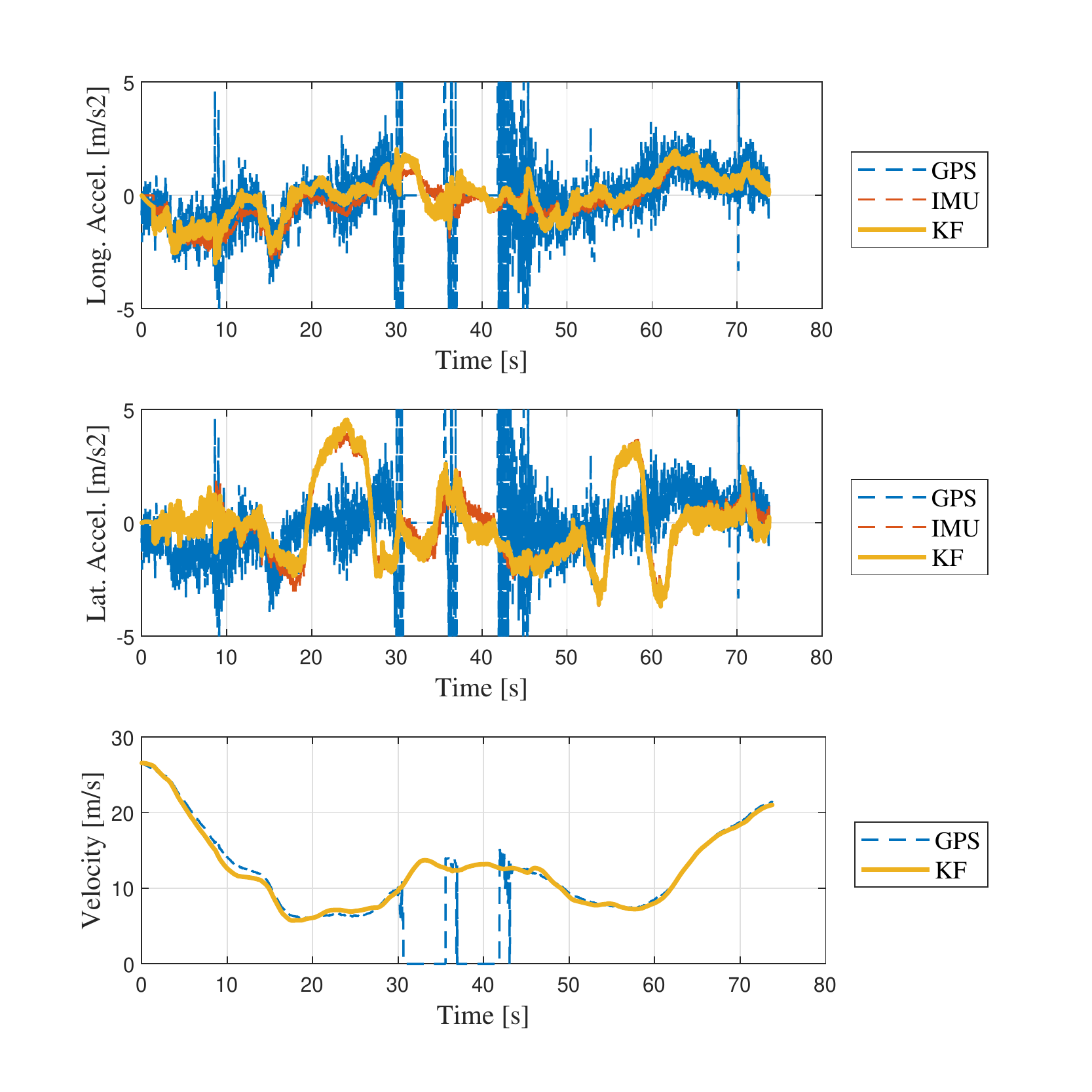}
    \caption{Raw (GPS and IMU) and processed (KF) data from one of the experimental test runs.}
    \label{fig:hd_process}
\end{figure}

\begin{figure}
    \centering
    \includegraphics[trim={20pt 0pt 40pt 10pt}, clip, width=\linewidth]{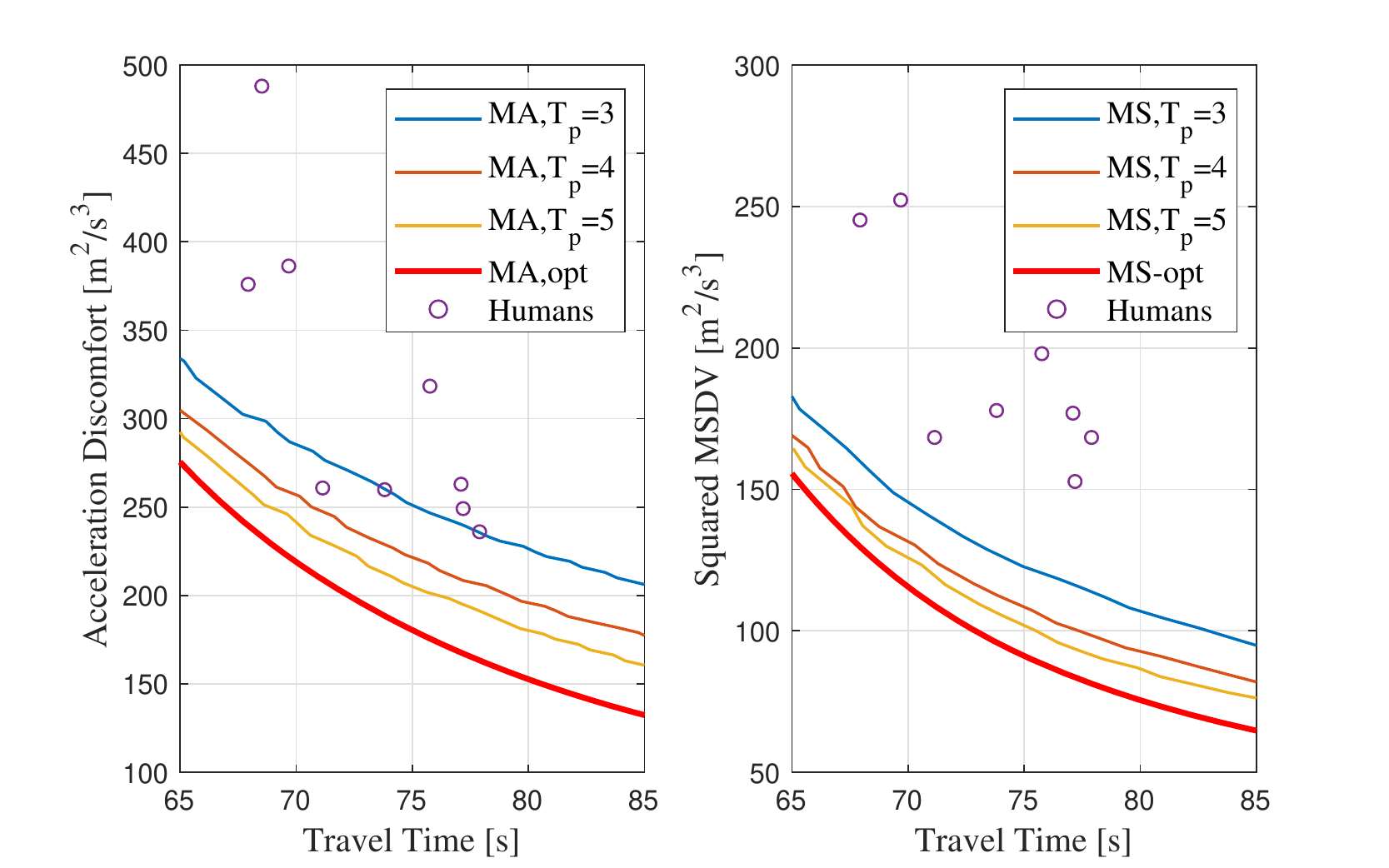}
    \caption{Distribution of the participants' driving performance in terms of motion sickness and acceleration comfort versus time efficiency.}
    \label{fig:hd_compare}
\end{figure}

\noindent An example of a test run performed by one of the volunteers is presented in Fig \ref{fig:hd_process}. The drive took \SI{73.8}{\second} long according to GPS data. The Kalman filter is able to comprehensively utilize both sources of information and reconstruct a more convincing motion profile. The total acceleration discomfort is \SI{259.8}{\square\meter\per\cubic\second} and squared MSDV is \SI{177.9}{\square\meter\per\cubic\second}. The group distribution of driving performance is shown in Fig. \ref{fig:hd_compare}. On some occasions, human drivers matched or slightly outperformed a receding-horizon motion planner with $T_p=$ \SI{3}{\second} in terms of general acceleration comfort. When it comes to motion sickness or frequency-weighted acceleration, however, human drivers performed worse than all variants of motion planners. In the best case, the deficiency is measured at approximately 20\% in squared MSDV. This might suggest that human drivers are not able to actively sense a difference in the frequency components in vehicle accelerations, or that subjective passenger comfort is different for drivers and passengers of a vehicle. The development of motion sickness has rather slow dynamics and the effects only become observable in the long term. Besides, the examples of human driving with very short travel time (under \SI{70}{\second}) have a higher deficiency in comfort and nauseogenicity. One may argue that human drivers are less experienced with driving in the highly dynamic range. They could find it difficult to follow the desired path accurately when given less time to observe and react.\par

\section{Conclusions}

\subsection{Findings}
\noindent This work proposed a novel formulation for incorporating the frequency sensitivity to accelerations in the development of a comfort-oriented optimization-based motion planning framework. The proposed method reduced the squared MSDV, an indicator of motion sickness, by up to 11.3\% compared with minimizing accelerations in general. A receding-horizon formulation of the optimal planning problem is compared with optimizing for a complex driving scenario in an integral manner. The performance difference from the integral planning approach is quantified for a variety of combinations of the preview parameters in the receding-horizon planner. We consider a preview time of \SI{3}{\second} as the lower limit since this leaves the vehicle with limited time to prepare for a corner entry, resulting in worse motion comfort and a higher chance of motion sickness. We found a longer nominal sampling time of \SI{0.5}{\second} beneficial for the purpose of mitigating motion sickness due to the slow dynamics in the frequency-weighting filter. On the other hand, a nominal sampling time of \SI{0.2}{\second} is sufficiently short for optimizing acceleration comfort. We further investigated the computation time of the receding-horizon planners. The algorithm is close to achieving real-time capability if it could be compiled as machine code and certain parameters of the optimization solver could be adjusted. Furthermore, on-road experiments revealed a performance gap between human drivers and the optimization-based planners described in this study. The proposed sickness-mitigating method achieved a reduction in frequency-weighted accelerations by 32\% from the best-performing human driver, while its counterpart without considering the frequency sensitivity improves general acceleration comfort by 19\%. The difference suggests that automated vehicles can be even more effective in mitigating motion sickness than in improving general motion comfort.

\subsection{Limitations and Future Works}
\noindent Our study relies primarily on the conclusion from past studies that motion sickness has a strong correlation with accelerations in certain frequency ranges. In the literature, there are more complex numerical models to predict motion sickness. We expect that including such models in the motion planning algorithm would require more intensive computation, making it more difficult to implement our approach in real-time. Alternatively, one may explore learning-based approaches to shift the computation offline and achieve real-time sub-optimal planning. Besides, the actual effect of the proposed method on the development of motion sickness among passengers in an automated vehicle remains to be validated. If an on-road experiment is to be performed, a closed track would be needed as motion sickness develops rather slowly and an instrumented vehicle needs to complete multiple laps to observe a difference. Lastly, this work includes only a small number of participants for the evaluation of human driving performance. This limits our ability to generalize the conclusions above to average human drivers. Instead, only the advantage over the best-performing participant was emphasized. We plan to test human drivers on a larger scale on public roads to solidify the baseline performance. This would contribute not only to the evaluation of motion planning and control methods in automated driving but also to the quantify their potential advantage over human drivers.\\


%





\ifCLASSOPTIONcaptionsoff
  \newpage
\fi



\bibliographystyle{IEEEtran}
\bibliography{references}
%



%

\begin{IEEEbiography}[{\includegraphics[width=1in,height=1.25in,clip,keepaspectratio]{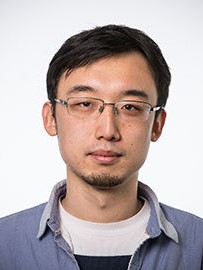}}]{Yanggu Zheng}
received the B.Sc. degree in Automotive Engineering from Tsinghua University, Beijing, China in 2015, and the M.Sc. degree (cum laude) from Delft University of Technology, Delft, the Netherlands in 2018. He is currently pursuing the Ph.D. in the Section of Intelligent Vehicles, Department of Cognitive Robotics, at Delft University of Technology. His research focuses on the application of optimization-based control and motion planning methods on automated vehicles for safety and comfort.
\end{IEEEbiography}

\begin{IEEEbiography}[{\includegraphics[width=1in,height=1.25in,clip,keepaspectratio]{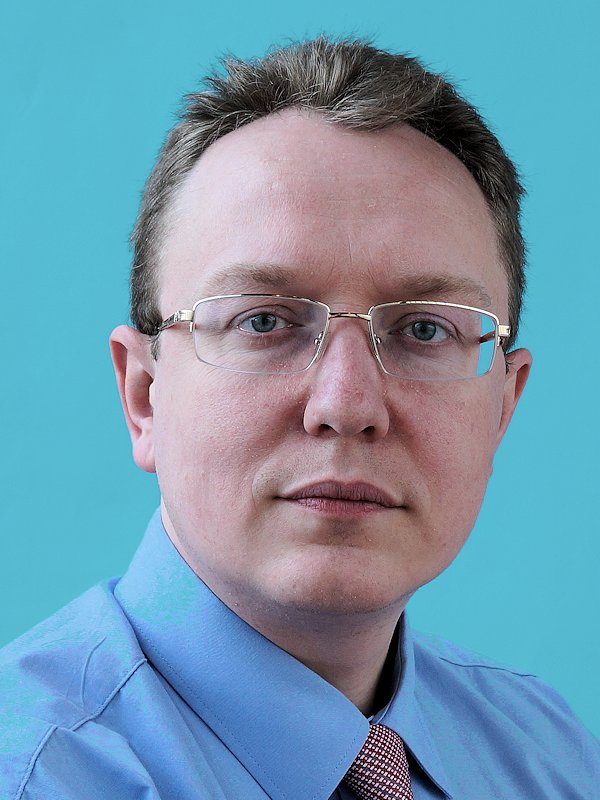}}]{Barys Shyrokau}
received the DiplEng degree (cum laude), 2004, in Mechanical Engineering from the Belarusian National Technical University, and the joint Ph.D. degree, 2015, in Control Engineering from Nanyang Technological University and Technical University Munich. He is an assistant professor in the Section of Intelligent Vehicles, Department of Cognitive Robotics, at Delft University of Technology and is involved in research related to vehicle dynamics and control, motion comfort, and driving simulator technology. He is a scholarship and award holder of FISITA, DAAD, SINGA, ISTVS, and CADLM.
\end{IEEEbiography}

\begin{IEEEbiography}[{\includegraphics[width=1in,height=1.25in,clip,keepaspectratio]{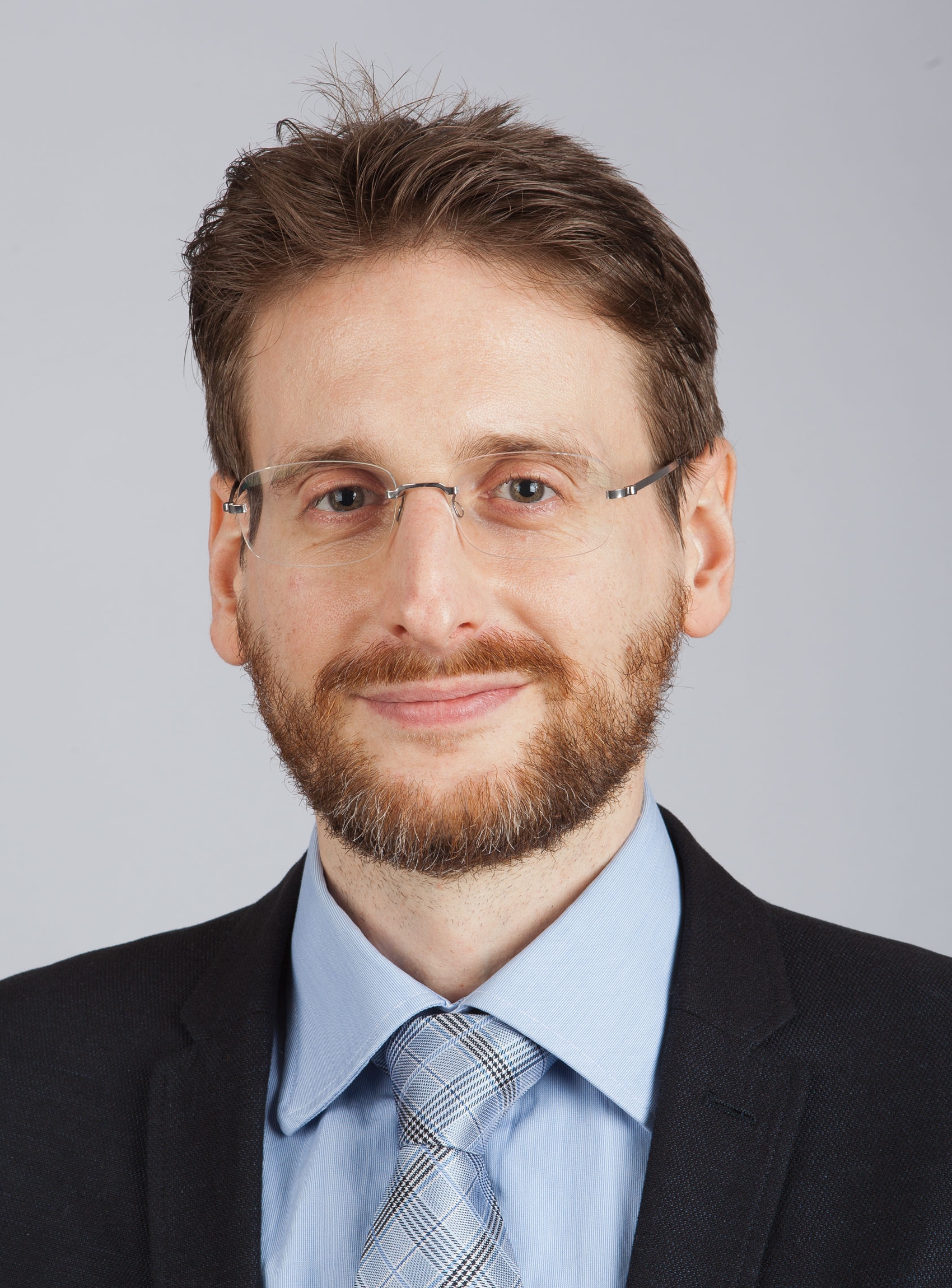}}]{Tamas Keviczky}
(Senior Member, IEEE) received the M.Sc. degree in electrical engineering from Budapest University of Technology and Economics, Budapest, Hungary, in 2001, and the Ph.D. degree from the Control Science and Dynamical Systems Center, University of Minnesota, Minneapolis, MN, USA, in 2005. He was a Post-Doctoral Scholar of control and dynamical systems with California Institute of Technology, Pasadena, CA, USA. He is currently a Professor with Delft Center for Systems and Control, Delft University of Technology, Delft, The Netherlands. His research interests include distributed optimization and optimal control, model predictive control, embedded optimization-based control and estimation of large-scale systems with applications in aerospace, automotive, mobile robotics, industrial processes, and infrastructure systems, such as water, heat, and power networks. He was a co-recipient of the AACC O. Hugo Schuck Best Paper Award for Practice in 2005. He has served as an Associate Editor for Automatica from 2011 to 2017 and for IEEE Transactions on Automatic Control since 2021.
\end{IEEEbiography}





\end{document}